# Ferromagnetism in LaFeO$_3$/LaNiO$_3$ Superlattices with High Curie Temperature

Tianlin Zhou[1,2], Fei Gao[1], Qinghua Zhang[1,2], Yuansha Chen[1,2], Xinzhe Hu[1], Yuzhou He[1], Yuchen Zhao[1,2], Jianjie Li[1,2], Minghang Li[1,2], Shaojin Qi[1,2], Fengxia Hu[1,2], Jirong Sun[1,2], Yunzhong Chen[1,2*,] and Baogen Shen[1-4]

[1]Beijing National Laboratory of Condensed Matter Physics and Institute of Physics, Chinese Academy of Sciences, Beijing 100190, China

[2]School of Physical Sciences, University of Chinese Academy of Sciences, Beijing 100049, China

[3]Ganjiang Innovation Academy, Chinese Academy of Sciences, Ganzhou, Jiangxi 341000, China

[4]CAS Key Laboratory of Magnetic Materials and Devices and Zhejiang Province Key Laboratory of Magnetic Materials and Application Technology, Ningbo Institute of Materials Technology and Engineering, Chinese Academy of Sciences, Ningbo 315201, China

Email: yzchen@iphy.ac.cn

**Interfacing complex oxides in atomically engineered layered structures can give rise to a wealth of exceptional electronic and magnetic properties that surpass those of the individual building blocks. Herein, we demonstrate a ferromagnetic spin order with a high Curie temperature of 608 K in superlattices consisting of otherwise paramagnetic perovskite LaNiO$_3$ (LNO) and antiferromagnetic LaFeO$_3$ (LFO). The extraordinary ferromagnetism likely results from the covalent exchange due to interfacial charge transfer from Fe to Ni cations. By deliberately controlling the thickness of the LNO sublayers thus the amount of charge transfer, a robust ferromagnetism of 4 $u_B$ is realized for a stacking periodicity consisting of one single unit cell of both LNO and LFO, an emergent double perovskite phase of La$_2$FeNiO$_6$ with B-site layered ordering configurations. The ferromagnetic**

**LFO/LNO superlattices offer great potential for the search of emergent magnetodielectric and/or multiferroic properties as well as applications in spintronics and electrocatalysts.**

**Introduction**

Atomically engineered heterostructures of strongly correlated oxides have enabled the rational design of emergent interfacial magnetic and electric properties not present in the constituting layer [1-3]. The most prominent examples are the two-dimensional (2D) electron gas formed at the interface between two band-gap insulators [4,5], superconducting interfaces between non-superconducting cuprates [6], and the ferromagnetism between two antiferromagnetic materials [7]. Among the magnetically ordered materials, including ferromagnetic (FM), ferrimagnetic, and antiferromagnetic (AFM), FM and ferrimagnetic materials have significant technological importance in diverse applications, such as permanent magnets, microwave devices, sensors, magnetic recording and memory media [8]. Magnetic interfaces particularly those based on perovskite oxides, $ABO_3$, are of similar technological importance, but the interface-induced changes to electronic structure, orbital occupancy, and charge redistribution in addition to metal-oxygen band hybridization often led to magnetic exchange interactions across B-O-B bonds different from those of the bulk [9-11]. Notably, although the charge transfer and the emerging functionalities at oxide heterointerfaces have been intensively investigated for early transition metal oxides [9, 11-14], the charge transfer across heterostructures consisting of two late transitional metal oxides remains less investigated [10, 15-17]. The research is even extremely rare when the thickness of the component for the heterointerface or the superlattice is precisely controlled down to a single unit cell (uc) of the perovskite oxide (approximately 0.4 nm) [18,19].

Rare-earth nickelate $R$NiO$_3$ (with $R$=rare-earth) is a typical family of late transition metal oxides, which have attracted significant interest in recent years due to their remarkable properties including sharp metal-to-insulator transitions (MIT) tunable with the $R$ radius [19-21] as well as the emergency of superconductivity in the infinitely layered $Nd_{0.8}Sr_{0.2}NiO_2$ thin films [22]. $LaNiO_3$ (LNO) is the only member of the rare earth perovskite $R$NiO$_3$ that is a Pauli paramagnetic (PM) metal with no signature of MIT or magnetic order at any temperature. Electronically, LNO is a charge transfer metal with strong hybridization between Ni 3$d$ and oxygen 2$p$ states, where the electronic configuration can be described as a superposition form $|\psi\rangle = \alpha|3d^7\rangle + \beta|3d^8\underline{L}\rangle$ ($\underline{L}$ denotes a ligand hole on the oxygen ion) [23]. The system shows partial covalence, and the degree of covalence is given by the ratio between $\beta^2$ and $\alpha^2$ ($\alpha^2 + \beta^2 = 1$). The hybridization between the 3$d$ band and the oxygen 2$p$ band results in the formation of oxygen holes and a small or negative charge transfer gap [21], which are key features of high-temperature superconducting cuprates [24]. Therefore, strain-engineered thin films or multilayers based on nickelates have been intensively investigated to explore electronic structures that would mimic those of cuprate parent materials [25], which prefer AFM order. Moreover, the low-lying Ni 3$d$ bands of LNO with partial filling make it prone to electron acceptors. Notably, ferromagnetic-like behavior has been reported in the paramagnetic LNO layer when it is interfaced with another ferromagnetic layer of $LaMnO_3$ thin films [15,16] or a ferrimagnetic layer of $GdTiO_3$[11]. But most of such charge-transfer-induced ferromagnetism shows Curie temperature below room temperature. In this Letter, we report a charge transfer induced ferromagnetism in the heterostructure of $LaFeO_3$/$LaNiO_3$ (LFO/LNO) (Fig.1a), where LFO is a G-type AFM insulator (Fig.1b) as electron donor, the amount of the transferred charge is controlled by the thickness of LNO on the scale of 1 uc. In the extreme limit of the single superlattice (SL) stacking periodicity (periodic lattice, p.l.) consisting of one unit cell (uc) pseudo-cubic LNO and one unit cell pseudo-cubic LFO, similar to the theoretically predicted double perovskite (DP) $La_2NiFeO_6$ [26], the SL shows not only a high magnetization of 4 $u_B$/p.l., but also a high Curie temperature of 608 K. The emergent properties are not observed either in the constituting materials (LFO and LNO) or the mixed compound $LaFe_{0.5}Ni_{0.5}O_3$ [27], which could result from the formation of an emergent DP phase of $La_2FeNiO_6$ with $B$-site layered ordering configurations along the (001) orientation. The ability to affect interfacial charge transfer and the Ni 3$d$-O 2$p$ covalency thus the Ni 3$d$-O 2$p$-Fe 3$d$ exchange interactions through a precisely controlled approach, not realizable by conventional processing techniques, suggests a new avenue to realizing extraordinary FM materials.

**Results**

**Creation of LNO$_n$/LFO$_1$ superlattices with layered ordering of cations**

We have grown LNO$_n$/LFO$_m$ superlattices (SLs) by alternately stacking LNO and LFO layers using pulsed laser deposition on SrTiO$_3$ (STO) (001) substrates (see Methods), where $n$ and $m$ represent the layer thickness of LNO and LFO in pseudo-cubic uc, respectively. Here $n$ was varied from $n = 1$ to 5 uc, while the LFO layer was set to 1 uc thick. The LNO$_n$/LFO$_1$ bilayer was repeated 10-20 times to form the SL. The SL growth was monitored *in-situ* by reflection high-energy electron diffraction (RHEED). Streaked RHEED patterns with clear intensity oscillations were observed during the growth of both LNO and LFO films, implying a 2D layer-by-layer film growth mode (Supplementary Figure S1). The epitaxial growth of the high quality LNO$_n$/LFO$_1$ SLs was also confirmed by X-ray diffraction experiments and high-resolution scanning transmission electron microscopy (STEM). Notably, although clear Laue fringes are visible for all the SLs from high-resolution XRD $\theta$–$2\theta$ scans (Supplementary Figure S2) due to their atomically smooth surfaces as confirmed by atomic force microscopy (Supplementary Figure S3), the satellite peaks indicating the chemically modulated structure with well-defined interfaces were only visible for $n \geq 3$, consistent with the previous results on similar structure [17]. Moreover, for $n \leq 3$, the SL is under compressive strain with $c \geq 3.905$ Å. While as $n>3$, the SL is under tensile strain. Such decrease of out of plane $c$ with increasing $n$ is consistent with the smaller lattice constant of LNO compared to LFO (LFO is orthorhombic with a pseudo-cubic lattice constant $a_0 = 3.93$ Å [17], and LNO is rhombohedral with pseudo-cubic $a_0 = 3.838$ Å). Figs.1c and d show the atomically resolved high-angle annular dark-field (HAADF) image of the cross-section of the LNO$_1$/LFO$_1$ SL and LNO$_4$/LFO$_1$ SL samples, respectively, recorded along the [100] zone axis. As the intensity of HAADF-STEM images is sensitive to atomic number (Z), both SLs show coherent epitaxial lattice as demonstrated by the brightest sublattice of the La ions. No dislocations, defects or any nano-scale phase-separated microstructures such as Fe or Ni metal clusters were observed. The compositional electron energy-loss spectroscopy (EELS) maps obtained from the analysis of the La M$_{4,5}$ (Fig.1e left), Fe L$_{2,3}$ (Fig.1e right), and Ni L$_{2,3}$ signals reveal the epitaxial growth of the periodic layered structure, which is flat and continuous over long lateral distances. Cation intermixing of Fe and Ni ions when occurs is found to be confined within 1-2 uc of LFO (within 1 p.l. of the SL). Shortly, high quality LNO$_n$/LFO$_1$ superlattices with layered ordering of configurations for Fe and Ni cations are successfully fabricated.

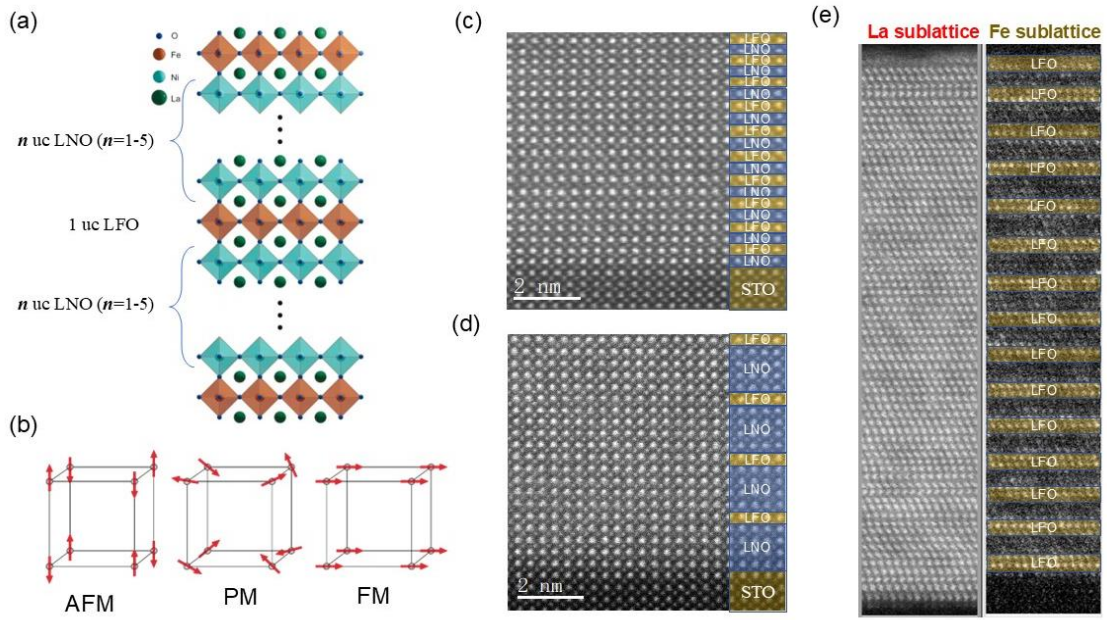

**Fig.1 Epitaxial growth of the LNO$_n$/LFO$_1$ SLs. a,** the sketch of the SL structure where the thickness of paramagnetic LNO, $n$, was varied from $n = 1$ to 5 uc, while the G-AFM LFO layer was set to 1 uc; **b,** the illustration for magnetic structures of G-AFM, paramagnetic (PM) and FM coupling; **c** and **d,** the cross-sectional STEM HAADF images of the LNO$_1$/LFO$_1$ and LNO$_4$/LFO$_1$ SL, respectively, recorded along the [100] zone axis. **e,** The EELS maps obtained from the analysis of the La M$_{4,5}$ (left)
5

and Fe $L_{2,3}$ (right) signals reveal the epitaxial growth of the periodic structure and cation intermixing confined within 1-2 uc of LFO (within 1 p.l. of the SL).

**Magnetic and transport properties of the emergent ferromagnetic superlattices**

LNO is a paramagnetic metal, and LFO is a G-type AFM insulator with intralayer antiparallel spin alignments, which shows one of the highest ordering temperatures of perovskite oxide ($T_N \approx 740$ K). Remarkably, when they were epitaxially grown along the (001) directions but absent for the (110) and (111) directions (Supplementary Figure S4), the $LNO_n/LFO_1$ SLs show strong signatures of ferromagnetism, which depends critically on the stacking thickness, $n$, of LNO. Fig.2a shows the hysteresis curves of the $LNO_n/LFO_1$ SLs measured at 300 K, $i.e.$ the dependence of the magnetization ($M$) as a function of the magnetic field $B$, which is saturated under the applied field above 3.5 kOe. All samples exhibited key characteristics of FM behavior, including hysteresis and remnant magnetization. At $n = 5$, the SL shows weak saturation ferromagnetism similar to $LaFe_{0.5}Ni_{0.5}O_3$ thin films. Remarkably, as $n$ decreases, the highest saturation magnetization increases significantly, reaching 252.8 emu/cm$^3$ for $n = 1$, approximately 4.0 $\mu_B$/p.l., comparable to the 4.96 $\mu_B$/f.u. (f.u.= formula unit) for the FM DP $La_2NiMnO_6$ [28], and much larger than that of the $LaFe_{0.5}Ni_{0.5}O_3$ thin films (Fig.2b). Fig.2c shows the temperature dependence of magnetization ($M$-$T$ curve) for the typical $LNO_n/LFO_1$ ($n=1$, 3 and 5) SL samples in addition to the $LaNi_{0.5}Fe_{0.5}O_3$ film. The Curie temperature, $T_C$, could be determined based on the minimum value of the temperature coefficient TCM, defined as TCM = 1/M(dM/dT). Remarkably, all the SL samples show a $T_C$ much higher than room temperature, around $T_C$=608 K, 589 K and 419 K, respectively, for $n$=1, 3 and 5. It is also noteworthy that the $LaNi_{0.5}Fe_{0.5}O_3$ film shows a $T_C$=109 K, much lower than those of the SL but higher than the bulk spin glassy temperature of 83 K [27]. The results strongly suggest that a unique FM order is accessed in the SLs structures with the strongest magnetism obtained at $n$=1, an emergent $La_2NiFeO_6$ DP structure with $B$-site layered ordering configuration along (001) direction. As illustrated in Table 1, its $T_C$ is close to the highest value ($T_C$ = 625

K) reported for $Sr_2CrReO_6$ among the DP compounds [29]. The temperature dependent resistivity, $\rho$, of the $LNO_n/LFO_1$ ($n$=1, 3 and 5) SL sample was also measured between 2-300 K, as shown in Fig.2d. All the SLs show a semiconductive behavior with small polaron conduction mechanism ($n \leqslant 4$) (Supplementary Figure S5), although the $LaFe_{0.5}Ni_{0.5}O_3$ thin film is found to be highly insulating. This strong conduction difference, in addition to the strong difference in magnetism, may be related to the difference in the ordered arrangement of the Fe and Ni cation atoms over the *B*-sites, where the higher level of cations ordering prefers higher magnetization and higher conductivity [19]. We thus obtained an emergent $La_2NiFeO_6$ DP with Curie temperature much higher than room temperature, which has been predicted to be a half-metallic ferromagnet [26].

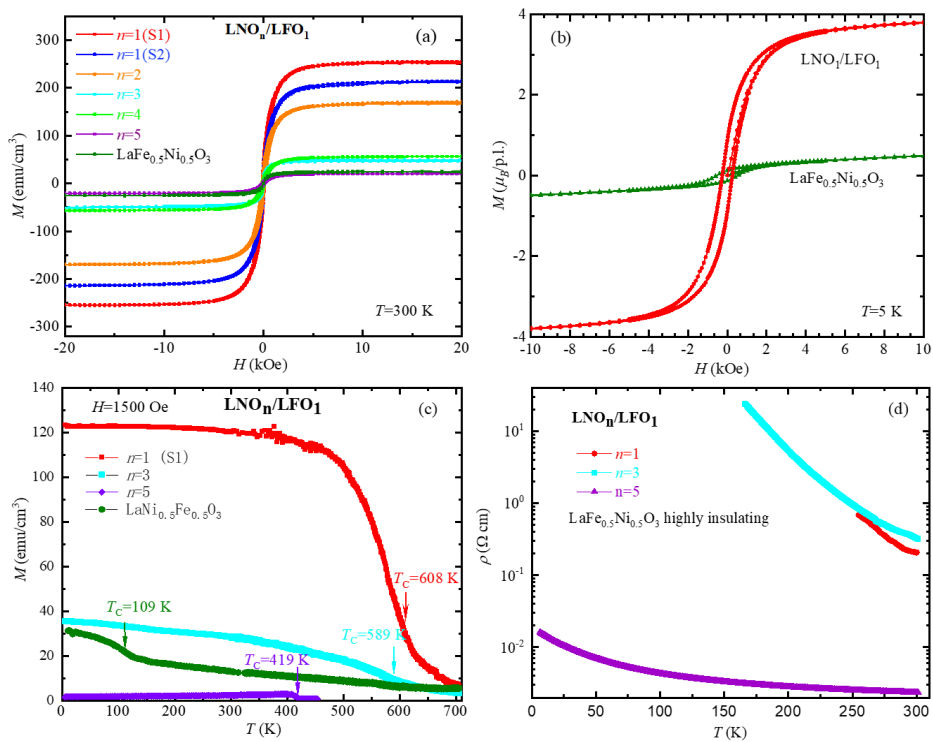

**Fig.2 Magnetic and transport properties of the atomically engineered $LNO_n/LFO_1$ SLs. a**, the hysteresis curves of the $LNO_n/LFO_1$ SLs measured at 300 K, for comparison the results of $LaFe_{0.5}Ni_{0.5}O_3$ thin films are also presented; **b**, the hysteresis curve of the $LNO_1/LFO_1$ SL measured at 5 K in comparison to the $LaFe_{0.5}Ni_{0.5}O_3$ thin film; **c**, the temperature dependence of magnetization for the $LNO_n/LFO_1$ ($n$=1, 3 and 5) SLs and the $LaNi_{0.5}Fe_{0.5}O_3$ film, measured at $H$=1500 Oe; **d**, the temperature dependent

resistivity, $\rho$, of the LNO$_n$/LFO$_1$ ($n$=1, 3 and 5) SLs, showing semiconductive behavior.

**Table 1 | The emerging ferromagnetic SLs and the typical double perovskites (DP) with ferromagnetism near and above room temperature**

| Formula | Structure | Saturation magnetization $M_s$ ($\mu_B$/p.l. or f.u.) | Curie Temperature $T_C$ (K) |
|---|---|---|---|
| LaFeO$_3$/LaCrO$_3$[7] | 1:1 SL | 3.0 | 375 |
| LaNiO$_3$/LaMnO$_3$[16] | 7:7 SL | 3.7. | 175 |
| **LaNiO$_3$/LaFeO$_3$** | **1:1 SL** | **4.0** | **608** |
| LaNiO$_3$/LaFeO$_3$ | 3:1 SL | 1.3 | 589 |
| La$_2$NiMnO$_6$ [28] | DP | 4.96 | 280 |
| Sr$_2$FeMoO$_6$[29] | DP | 3.98 | 345 |
| Sr$_2$CrReO$_6$[29] | DP | 2.6 | 625 |

**Magnetic domain microstructure characterized by magnetic force microscopy**

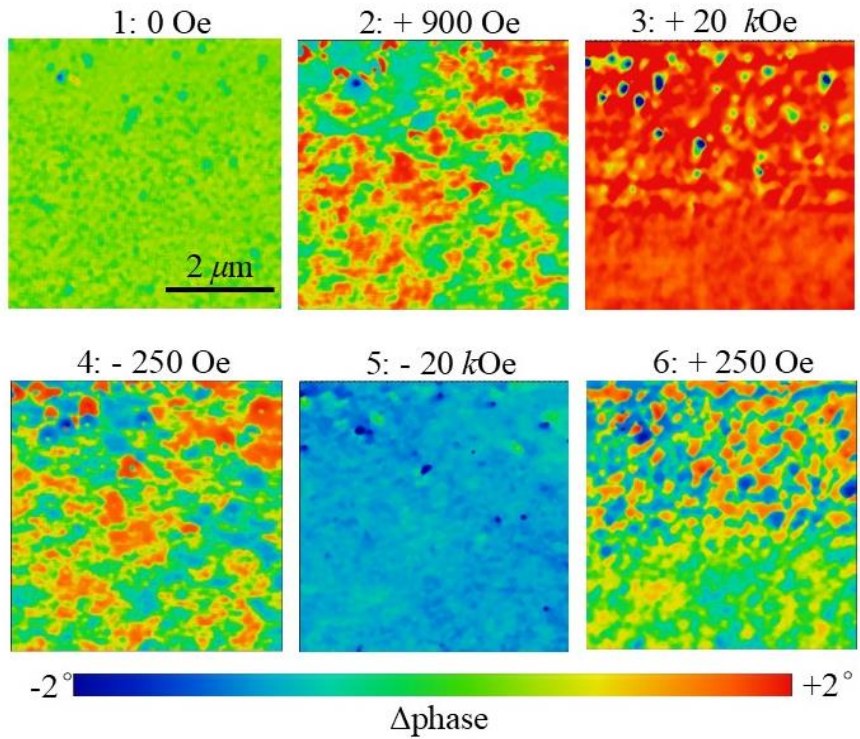

**Fig.3 Magnetic force microscopy (MFM) domains structure of the LNO$_1$/LFO$_1$**

**SL sample obtained at 2 K.** The magnetic contrasts ($\Delta\varphi$) are represented by colors: the red (blue) colored area with positive (negative) $\Delta\varphi$ represents the net magnetization projecting along the upward (downward) z direction, and the green colored area with nearly zero $\Delta\varphi$ represents for domain walls or the grains with zero-magnetization.

The magnetic domain microstructure and its evolution with respect to the applied field for the $LNO_1/LFO_1$ SL were also investigated by magnetic force microscopy (MFM). Generally, the magnetograms obtained by MFM provide information of the effective magnetization along the z direction of the local domains. Fig.3 shows the magnetograms obtained at 2 K in different out-of-plane magnetic fields corresponding to the critical points on the *M-H* loop in Fig.2b. Here the red or blue-colored regions represent net-magnetization oriented up ($+M_z$) or down ($-M_z$), respectively, and the green color represents the zero $M_z$ signal. As shown in Fig.3, the as-grown $LNO_1/LFO_1$ SL first demonstrates nearly zero-magnetization along the z direction after the zero-field-cooling (ZFC) process. By applying a positive magnetic field along the out-of-plane direction, discrete upward domains begin to appear and grow continuously (+ 900 Oe), eventually reaching an upward single-domain state at the saturation field of +20 *k*Oe. By applying a negative magnetic field, the single-domain state turns to fragments of upward and downward domains at the coercive field of approximately -250 Oe, corresponding to the zero net magnetization in the M-H loop. The typical size of the fragment domains after magnetization (measured at +250 Oe) is about 300~500 nm (Supplementary Figure S6). The magnetic domain evolution is repeatable during the field sweeping process, as further evidenced by the downward single-domain state at -20 *k*Oe and then the fragment-domain state at +250 Oe (the positive coercive field). The domain structures after ZFC and their evolution upon applying magnetic fields are consistent with the direct magnetization measurements. This further confirms that the strong magnetism is intrinsic to our SL samples, where the plausible magnetic contribution from the STO bulk substrate is negligible.

**Discussion**

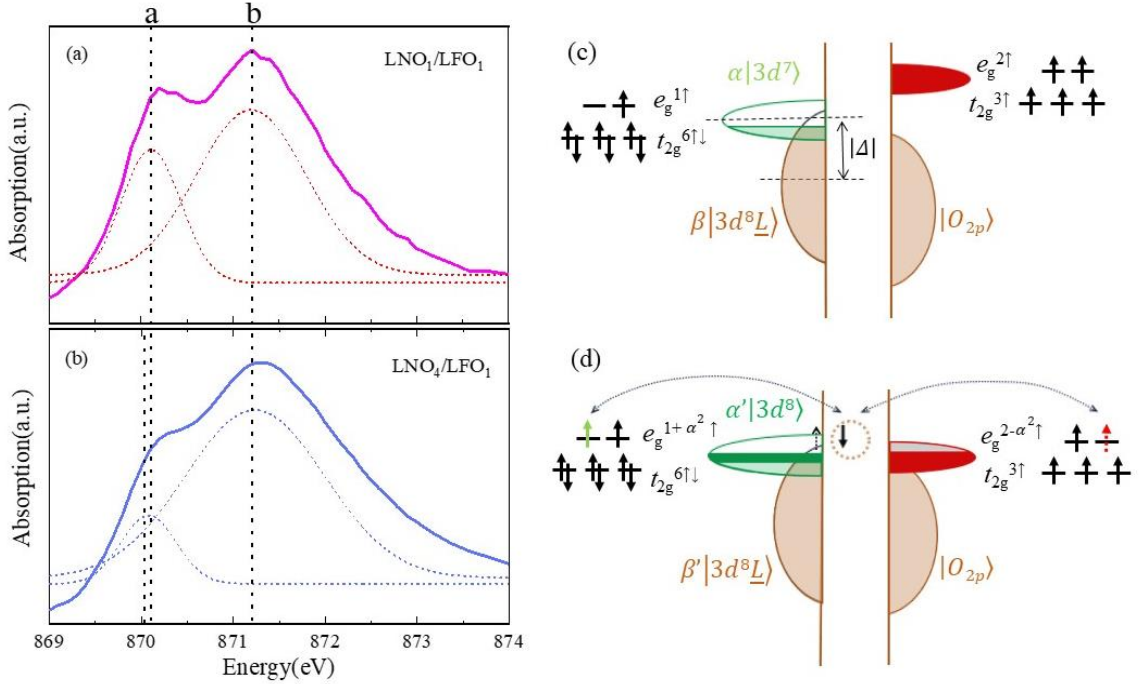

**Fig.4 Charge transfer at the $LNO_n/LFO_1$ SL and a schematic representation of the covalent ferromagnetism in the $LNO_1/LFO_1$ SL.** (a) and (b) The XAS signal of $LNO_1/LFO_1$ and $LNO_4/LFO_1$ near the Ni $L_2$ edge, respectively. All curves were fitted with two Gaussian functions, labeled as "a" and "b". The fitting data were shown as thin dashed lines. (c) and (d) the schematic representation of the charge state before and after the charge transfer induced rehybridization process, respectively, from LFO to LNO.

The magnetic exchange interactions in $3d$ perovskite oxides occur through $d$ orbitals (dominated mainly by $e_g$ orbitals) centered at neighboring $B$ cations bridged by an oxide ion, referred to as "super-exchange interaction". The guidelines dubbed as the Goodenough-Kanamori-Anderson rules can be used to predict the ferromagnetic and antiferromagnetic states for transition-metal oxides [30]. The nominal $Fe^{3+}$ ion in LFO has a $3d^5$ ($t_{2g}^{3\uparrow} e_g^{2\uparrow}$) configuration. It is AFM in accordance with the Goodenough-Kanamori rules. The nominal $Ni^{3+}$ ion has a $3d^7$ ($t_{2g}^{6\uparrow\downarrow} e_g^{1\uparrow}$) configuration with filled $t_{2g}$ orbitals and one electron residing in the doubly degenerate $e_g$ orbitals. Due to the electron-acceptor nature of LNO as illustrated in Fig.4c, the strong ferromagnetism could originate from the electron transfer from the LFO to the LNO layer, which fills not only $\alpha|3d^7\rangle$ but also $\beta|3d^8\underline{L}\rangle$ states, where the ferromagnetism at the interface could emerge from the covalent exchange coupling between $Ni^{2+}$ and $Fe^{4+}$ ions through the hybridization between the oxygen $2p$ band and the $3d$ band [31].

To determine the cation valence ($Fe^{3+}/Fe^{4+}$, $Ni^{3+}/Ni^{2+}$) thus the charge transfer in addition to their spatial distribution of the Fe-Ni system [32], X-ray absorption spectroscopy (XAS) measurements were performed on our $LNO_n/LFO_1$ ($n$=1 and 4) SL samples. The signal of the Fe $L_{2,3}$ edge was found to show roughly similar features between $LNO_1/LFO_1$ and $LNO_4/LFO_1$ samples, except that the splitting between the two Fe $L_3$ peaks of the $LNO_1/LFO_1$ was slightly wider than that of the $LNO_4/LFO_1$ sample (Supplementary Figure S7). In contrast, the XAS signal at the Ni $L_{2,3}$ edge, after correction for the overlap of the La $M_{4,5}$ signal, shows a distinct difference between $LNO_1/LFO_1$ and $LNO_4/LFO_1$ (Supplementary Figure S7). The shape of the spectra in our SL samples clearly deviates from that of the LNO film [33], which readily indicates a different formal Ni oxidation state. Particularly, as shown in Figs. 4a and b, we compared the Ni $L_2$-edge region for the $LNO_1/LFO_1$ and $LNO_4/LFO_1$ samples. This region has no overlap with La edges and shows quite distinct features between $Ni^{3+}$ and $Ni^{2+}$ thus can gain insight into the Ni valence in our samples as reported previously [11]. The Ni $L_2$ experimental spectra for the $LNO_1/LFO_1$ and $LNO_4/LFO_1$ samples were fitted with two Gaussian peaks labeled as 'a' and 'b'. The changes in the spectral shape, particularly the sharpening of the low-energy feature (labeled 'a'), are reminiscent of the signature of $Ni^{2+}$ in the NiO [34]. By comparing the relative intensity of the two features ('a' and 'b'), it is clear that the Ni valence is stronger reduced towards 2+ for the $LNO_1/LFO_1$ sample than the $LNO_4/LFO_1$ sample. These results not only confirm the charge transfer from Fe to Ni ions in the $LNO_n/LFO_1$ SL samples, but also reveal a strong dependence of the amount of transferred electrons on the thickness of the nickelate, where the charge transfer length is in a short range nature and strongly confined to the proximity of the interface.

Upon charge transfer, the filling of the $Ni^{3+}$ $\alpha|3d^7\rangle$ will increase towards $Ni^{2+}$ $\alpha'|3d^8\rangle$ and the level of the corresponding subband will decrease, thus the overlapping or the hybridization between the 3$d$ band and the oxygen 2$p$ band $\beta'|3d^8\underline{L}\rangle$ will also increase, as illustrated in Fig.4d. This rehybridization process requires a redistribution of electrons between the $|3d^7\rangle$ and the $|3d^8\underline{L}\rangle$ states, which occurs at an energy cost $|\Delta|$ for each electron transferred between $|3d^8\underline{L}\rangle$ and $\alpha|3d^7\rangle$, but thanks to the presence of $\alpha'|3d^8\rangle$, the $|\Delta|$ is expected to be decreased after change transfer and the rehybridization is more prone to occur. On the other hand, due to the strong covalent character of LNO with a large rare-earth size, which increases the cost of rehybridization, the competition between the charge transfer, controlled by the energy gain associated with the difference in electron affinity, and the cost of rehybridization, limits the amount of electrons that can be transferred across the interface. Notably, changes in the multiplet splitting of the Ni $L_3$ absorption edge can be used to estimate the level of covalence [35]. This splitting corresponds to the energy separation between $t_{2g}$ and $e_g$ levels resulting from the interplay between hybridization and Coulomb repulsion, which are both stronger for $e_g$ levels. Previous studies have shown that the larger splitting of the $L_3$ peaks indicates a decrease in covalence [11,34]. The fitted XAS at the Ni $L_3$ edge also indicates that peak splitting increases from $LNO_1/LFO_1$ to the $LNO_4/LFO_1$ samples, reflecting a decrease in covalence, where a weakened ferromagnetism is also observed. These results suggest that the charge transfer from $Fe^{3+}$ to $Ni^{3+}$ indeed results in an increase of covalence, favoring the ferromagnetic $Ni^{2+}$-$O^{2-}$-$Fe^{4+}$ interactions [31]. The hybridization and charge transfer along the interfacial Fe-O-Ni bonds are mainly proposed to explain the ferromagnetism across the interface and on the Ni site. As for the magnetic nature of the LFO single layer, in contrast to the bulk G-AFM configuration, it probably exhibits a helicoidal magnetic (HM) ordering

due to the negative charge transfer energy and the enhanced oxygen-oxygen hopping amplitude when it is interfaced with LNO upon charge transfer [36]. It is noteworthy that X-ray magnetic circular dichroism (XMCD) measurements showed negligible local Fe or Ni moments at the Fe $L_{2,3}$ and Ni $L_{2,3}$, although they were found to be coupled ferromagnetically to each other. As for the fact that the estimated Fe and Ni moment from XMCD is significantly lower than that measured by the SQUID magnetometry, it might result from the presence of a surface layer with a strongly reduced FM signal as the XMCD measurement was performed in total electron yield configuration, which is very surface sensitive and probes only a few nano-meter deep. Similar huge deviations have been also observed in $BaFeO_3$ thin films [37] and for $LaMnO_3/LaNiO_3$ superlattices [16]. Therefore, the exact origin of the strong ferromagnetism remains under investigation.

**Conclusions**

Shortly, robust ferromagnetism with high magnetization and Curie temperature much above room temperature was achieved in $LNO_n/LFO_1$ SL by precisely controlling the charge transfer as well as the interfacial orbital hybridization in late transitional metal oxides. The latter is viable for tuning the strength of the exchange interactions in all-oxide heterostructures. The semiconducting and FM $LNO_1/LFO_1$ SL, which can be regarded as an emergent DP phase of $La_2FeNiO_6$ with *B*-site layered ordering configurations along the (001) orientation, can have a wide spectrum of applications for spintronics.

**Methods:**

**Sample fabrication and characterization**:

High quality $LNO_n/LFO_1$ SLs were epitaxially grown on (001)-oriented STO substrates by pulsed laser deposition (KrF, $\lambda$=248 nm) with 10-20 repetitions. During film growth, the substrate temperature was kept at 620°C and the oxygen pressure was set to $3.4\times10^{-2}$ mbar. The fluence of the laser pulse was 1.2 J/cm$^2$, and the repetition rate was 2 Hz. The SL growth was monitored *in-situ* by reflection high-energy electron diffraction (RHEED). Layer by layer epitaxial growth was achieved for both the LNO and LFO layers. After deposition, the sample was cooled to room temperature under the oxygen pressure of 0.1 mbar. The crystal structure of the superlattice was observed by X-ray diffraction (XRD).

The magnetic properties were measured by a Quantum-Designed vibrating sample magnetometer (VSM-SQUID). The *M-T* measurements were performed in two temperature ranges from 2-300 K and 300 K-750 K using the same system with an oven option. The transport measurement was measured by a physical properties measurement system (PPMS) at temperatures from 5-300 K.

MFM measurements were performed in a variable temperature system from attocube equipped with a superconducting magnet (attoDRY2100). All measurements were conducted in vacuum at 2 K, based on a phase modulation technique in noncontact mode, with a cantilever (PPP-MFMR) with a spring constant k~2.8 N/m and a resonant frequency f~75 kHz.

**Scanning transmission electron microscopy (STEM):** For cross-sectional microscopy, the sample was prepared by using focused ion beam (FIB) milling. Cross-sectional lamellas were thinned down to 60 nm thick at an accelerating voltage of 30 kV with a decreasing current from the maximum 2.5 nA, followed by fine polish at an accelerating voltage of 2 kV with a small current of 40 pA. STEM measurements were conducted by a double Cs-corrected JEOL JEM-ARM200CF operated at 200 kV with a CEOS Cs corrector (CEOS GmbH, Heidelberg, Germany).

**X-ray absorption spectroscopy (XAS):** Resonant X-ray reflectometry (RXR) experiments were carried out at the REIXS beamline of the Canadian Light Source at 300 K in an ultrahigh vacuum environment.

**Acknowledgements**

The discussions with R. Green and C. Piamonteze are greatly appreciated. The authors thank the support from the Science Center of the National Science Foundation of China (52088101), the National Key Research and Development Program of China (2023YFA1406400, 2021YFA1400300), the National Natural Science Foundation of China (51327806, 52072400, 52322212, T2394472, T2394470), and the support from the Synergetic Extreme Condition User Facility (SECUF). Part or all of the research described in this paper was performed at the Canadian Light Source, a national research facility of the University of Saskatchewan, which is supported by the Canada Foundation for Innovation (CFI), the Natural Sciences and Engineering Research Council (NSERC), the Canadian Institutes of Health Research (CIHR), the Government of Saskatchewan, and the University of Saskatchewan.

**Author contributions**

Y.Z.C. designed the concept and experiments. T.L.Z, F.G, S.J.Q fabricated the samples and performed magneto-transport characterization. Y.S.C performed the MFM experiment and data analysis. H.Y.Z, Q.H. Z performed the STEM measurements and analysis. All authors discussed the results, interpretations, and wrote the manuscript.

**Competing interests:** The authors declare that they have no competing interests.

**Data and materials availability:** All data needed to evaluate the conclusions in the paper are present in the paper and/or the Supplementary Materials. Additional data related to this paper may be requested from the authors.